\documentstyle[twoside,fleqn,espcrc2]{article}

\newcommand{\AmS}{{\protect\the\textfont2
  A\kern-.1667em\lower.5ex\hbox{M}\kern-.125emS}}

\title{Partially quenched QCD and staggered fermions}

\author{Claude Bernard and Maarten Golterman\address{Department of Physics,
        Washington University\\
        St. Louis, MO 63130-4899, USA}%
}

\begin{document}

\begin{abstract}
We summarize results for partially quenched chiral perturbation theory
and indicate an application to staggered fermion QCD in which the square
root of the determinant is taken to reduce the number of flavors from four
to two.
\end{abstract}

\maketitle

\section{Partially quenched chiral perturbation theory}

We summarize the results of a recent investigation of
partially quenched theories
\cite{uspq}.  This extends previous work on the fully quenched case
\cite{usPRD,uslatams} (see also refs. \cite{sharpeone,sharpetwo}).
Partially quenched theories are theories
in which not all fermions are quenched; only for $k$ of the $n$ fermions
present in the theory will the determinant in the functional integral
be replaced by $1$.  A lagrangian formulation of such a theory is obtained
by considering QCD with $n+k$ quarks, where the first $n$ quarks are just
normal quarks (denoted by $q_i$, $i=1\dots n$), whereas the other $k$
quarks (denoted by ${\tilde q}_j$, $j=1\dots k$) will be given bosonic
statistics.
We will refer to such a theory as an $SU(n|k)$ theory.
The same method that we developed for
studying ChPT for a completely quenched theory can also be applied in
this case.

Our motivation for considering partially quenched theories
is threefold:

First, one may learn more about the peculiar infrared
behavior \cite{usPRD,uslatams,sharpetwo}
by considering what happens when only part of the fermion
content of a theory is quenched.
In particular, if different fermion mass scales
are present, one might ask how the
infrared behavior depends on whether all or only some of the fermions
with a common mass are quenched.  Also, it is
interesting to know what
happens in the unquenched sector of the theory: is a
theory with $n$ fermions, out of which $k$ are quenched, the same as an
unquenched theory with just $n-k$ fermions?

Second,
partially quenched theories arise naturally in the
description of simulations in which the valence quark masses are not
chosen  equal to the sea-quark masses.  This is a not uncommon
numerical technique which, for example, allows one to use
Wilson valence quarks and staggered  sea-quarks.  One would
like to have a chiral theory for such simulations.

A third motivation comes from staggered fermion QCD, which
describes QCD with four
flavors of quarks in the continuum limit.  In order to use these
fermions for simulations of QCD with only two flavors, a common
trick
is to take the square root of the fermion
determinant, thereby effectively reducing the number of flavors which
appear in virtual quark loops from four to two.

We can state three theorems about partially
quenched theories:

\begin{description}
\item[I.] In the subsector where all valence quarks
are unquenched
the $SU(n|k)$ theory is completely
equivalent to a normal, completely unquenched $SU(n-k)$ theory.
\item[II.] The ``super-$\eta'$,'' $\Phi_0$, defined as
\begin{equation}
\Phi_0 = \frac{1}{\sqrt{n-k}}(\sum_{i=1}^n \bar q_i \gamma_5 q_i +
\sum_{j=1}^k \bar{\tilde q}_j \gamma_5 \tilde q_j)\ ,
\end{equation}
is equivalent to the $\eta'$ constructed in
the unquenched sector of the $SU(n|k)$ theory, and is therefore,
by I, equivalent to the $SU(n-k)$ $\eta'$.  ``Equivalent''
here means that Green's functions constructed from an arbitrary
number of super-$\eta'$ fields and unquenched quarks, will be equal to
the corresponding Green's functions with the super-$\eta'$ replaced
by the $\eta'$ of the
$SU(n-k)$ theory.
\item[III.] Quenched infrared divergences,
 coming from a double
pole in the $\eta'$ propagator and associated with some quark
mass of mass $m$, will arise if and only if the scale
$m$ is fully quenched, {\it i.e.}, if there is a pseudoquark of mass $m$
for every quark of mass $m$.
\end{description}

Theorems I and II can be proved by simple physical arguments, which
rely on the cancellation between quarks and pseudoquarks
(bosonic quarks) in virtual loops (Thrm.\ I) or valence lines
(Thrm.\ II).  Theorem III requires a detailed examination of the
propagator in the neutral
meson sector in partially quenched chiral perturbation theory.
It is then possible to show
the offending double poles can only arise when
a mass scale is fully quenched \cite{uspq}.

\section{Staggered fermions}

In this section, we will consider the definition of two-flavor meson
operators in the mass degenerate two-flavor theory obtained from the
degenerate four-flavor theory in which the square root of the
determinant is taken.  Theorem I tells us that we can obtain
the two-flavor unquenched theory in this way, and that
no problems are to be expected from taking
the square root.  For nonsinglet mesons no tuning of the operators
is required because one may use the same operators as in the four-flavor
theory.

However, one expects that the definition of an operator for the
$\eta'_{SU(2)}$ in the four-flavor theory will require tuning
\cite{golsmi}, even with
degenerate quark masses.
What we wish to show here is
that nevertheless two ways exist for choosing a mass matrix and a meson
operator which do not require tuning of the operator in order to define
a pure $\eta'_{SU(2)}$ in the four-flavor theory.  The first method
consists of applying theorem II, whereas the second method makes use of
a peculiarity of nonlocal staggered fermion mass terms.

For staggered fermion QCD, mass terms can be constructed which lead
to the most general four flavor mass matrix $M$
in the continuum limit \cite{golsmi}:
\begin{eqnarray}
M & = & m+m_\mu\xi_\mu+\frac{1}{2}m_{\mu\nu}(-i\xi_\mu\xi_\nu) \nonumber \\
& & \qquad \mbox{} +m^5_\mu i\xi_\mu\xi_5+m^5\xi_5.
\end{eqnarray}
The $4\times4$ $\xi$-matrices form a representation of the Clifford
algebra $\xi_\mu\xi_\nu+\xi_\nu\xi_\mu=2\delta_{\mu\nu}$, and are
identified with $SU(4)$ flavor generators
in the continuum limit.  We will denote
the terms in eq. (2)
with scalar (S), vector (V), tensor (T), axial vector
(A) and
pseudoscalar (P) respectively.
They correspond to $0,\dots,4$ link operators in the staggered
fermion action.

It can be shown that this form of the mass matrix is stable under
renormalization, in the sense that the coefficients $m$, $m_\mu$,
$\dots$ will only receive multiplicative renormalizations, one for each
tensor structure in eq. (2) \cite{golsmi}.
Note that
the mass matrix $M$ needs to be  diagonalized in order to
determine what the mass eigenstates are.

Let us first consider the simplest possible mass matrix, by choosing
only the single site mass $m$ to be nonzero,
corresponding to four degenerate flavors.  In this case, the simplest
operator for an $\eta'_{SU(4)}$ will be a four link operator,
which in the continuum limit corresponds to the operator
$\bar{\psi}\gamma_5\psi$,
where $\psi$ is a continuum Dirac field with four
flavor components.

In this basis, an $\eta'_{SU(2)}$ would be created by the continuum
operator
\begin{equation}
\eta'^{\rm cont}_{SU(2)}=\bar{\psi}
\pmatrix{1&0&0&0\cr 0&1&0&0\cr 0&0&0&0\cr 0&0&0&0\cr}\gamma_5\psi.
\end{equation}
Clearly, in order to construct a staggered operator with this continuum
limit, we need an operator with flavor matrix
of type S to get a nonzero trace because the
$\eta'_{SU(2)}$ flavor matrix in
eq. (3) has a nonvanishing trace, and V, T, A and P are all
traceless.  In addition,
we need an operator of the type V, T, A
or P, since the matrix contains
two zero eigenvalues.  The fact that these operators renormalize
differently from S leads to the need to tune their relative coefficient.
We conclude that with a single
site mass term no explicit $\eta'_{SU(2)}$
operator can be constructed in
the four-flavor staggered theory without tuning.  The only way to avoid
tuning in this case, is to compute the diagrams for the $\eta'_{SU(4)}$,
and adjust the relative coefficients of the
straight-through and the two-hairpin diagrams,
as implied by theorem II, and explained in detail in
section 2 of ref. \cite{uspq}.

Actually, the special properties of the tensor operator make it possible
to construct an $\eta'_{SU(2)}$ without tuning in a different way.
To discuss this, we will choose an explicit representation of the
$\xi$-matrices:
\begin{equation}
\xi_i=\sigma_i\otimes\tau_1,\ \ \
\xi_4=\tau_2,\ \ \
\xi_5=\tau_3.
\end{equation}

In this case,
it is necessary to choose a mass term of the tensor type. For
definiteness we choose
\begin{equation}
M_0=m(-i\xi_1\xi_2)=\pmatrix{m&0&0&0\cr 0&-m&0&0\cr
0&0&m&0\cr 0&0&0&-m\cr},
\end{equation}
which corresponds again to four flavors with a degenerate mass $m$.
The minus signs can be removed by a nonanomalous chiral transformation.
The $\eta'_{SU(4)}$ with this mass matrix is
\begin{eqnarray}
\eta'^{\rm cont}_{SU(4)} & \propto &
\bar{\psi}(-i\xi_1\xi_2)\gamma_5\psi \nonumber \\
& = & \bar{\psi}
\pmatrix{1&0&0&0\cr 0&-1&0&0\cr 0&0&1&0\cr 0&0&0&-1\cr}\gamma_5\psi\ .
\end{eqnarray}
Projecting to $SU(2)$, we get for the $\eta'_{SU(2)}$
\begin{eqnarray}
\eta'^{\rm cont}_{SU(2)} & \propto &
\bar{\psi}
\pmatrix{1&0&0&0\cr 0&-1&0&0\cr 0&0&0&0\cr 0&0&0&0\cr}\gamma_5\psi
\nonumber \\
& =& \bar{\psi}(-i\xi_1\xi_2-i\xi_3\xi_4)\gamma_5\psi.
\end{eqnarray}
Unlike the previous case, this $\eta'_{SU(2)}$ flavor matrix is now
traceless, which allows us to write it as a sum of two tensor terms.

The $\eta'_{SU(2)}$
of eq. (7) is now constructed from two tensor
operators
rather than one scalar and one of some other type.  Since all tensor
operators get renormalized in the same way, no tuning is needed here.
The price, however, is the use of a tensor mass term, which would
make this approach awkward for standard simulations.
Using the
$\eta'_{SU(4)}$, and readjusting the relative weight of the diagrams by hand,
will
be preferable in most cases.  We note that such readjustment
is standard
practice in weak matrix element calculations with
staggered fermions \cite{kilcup}.
\vskip0.2in
{\bf Acknowledgements}
\vskip0.1in

We are greatful to Doug Toussaint for
inspiring this
project by
suggesting that our chiral techniques
could be used to examine the determinant square-root method
for staggered fermions.  We also thank Carl Bender and
Mike Ogilvie for very
useful discussions.
Part of this work was carried out at Los Alamos National Laboratory and
UC Santa Barbara.  M.G. would
like to thank Rajan Gupta and the Theory Division of LANL,
and both of us would like to thank the UCSB Physics Department,
and in particular Bob Sugar, for hospitality.

C.B. and M.G. are supported in part by the Department of Energy under grant
\#DOE-2FG02-91ER40628.

\end{document}